\documentclass[fleqn,usenatbib]{mnras}

\usepackage[T1]{fontenc}

\DeclareRobustCommand{\VAN}[3]{#2}
\let\VANthebibliography\thebibliography
\def\thebibliography{\DeclareRobustCommand{\VAN}[3]{##3}\VANthebibliography}

\usepackage{graphicx}	
\usepackage{amsmath}	
\usepackage{amssymb}	
\usepackage{newtxtext,newtxmath}

\title[Probing the shape of dust emission in local galaxies]{Probing the
  spectral shape of dust emission with the DustPedia galaxy sample}

\author[A. Nersesian et al.]{
Angelos Nersesian,$^{1, 2}$\thanks{E-mail: angelos.nersesian@ugent.be}
Wouter Dobbels,$^{1}$
Manolis E. Xilouris,$^2$
Maarten Baes,$^{1}$
\newauthor
Simone Bianchi,$^{3}$
Viviana Casasola,$^{4}$
Christopher J. R. Clark,$^{5}$
Ilse De Looze,$^{1, 6}$
\newauthor
Fr\'{e}d\'{e}ric Galliano,$^{7}$
Suzanne C. Madden,$^{7}$
Aleksandr V. Mosenkov,$^{8}$
\newauthor
Evangelos-D. Paspaliaris,$^{2, 9}$
and Ana Tr\v{c}ka$^{1}$
\\
$^{1}$Sterrenkundig Observatorium Universiteit Gent, Krijgslaan 281 S9, B-9000 Gent, Belgium\\
$^{2}$National Observatory of Athens, IAASARS, Ioannou Metaxa and Vasileos Pavlou GR-15236, Athens, Greece\\
$^{3}$INAF - Osservatorio Astrofisico di Arcetri, Largo E. Fermi 5,I-50125, Florence, Italy\\
$^{4}$INAF - Istituto di Radioastronomia, Via P. Gobetti 101, 4019, Bologna, Italy\\
$^{5}$Space Telescope Science Institute, 3700 San Martin Drive, Baltimore, Maryland, 21218, USA\\
$^{6}$Department of Physics and Astronomy, University College London, Gower Street, London WC1E 6BT, UK\\
$^{7}$Laboratoire AIM, CEA/DSM - CNRS - Universit\'{e} Paris Diderot, IRFU/Service d’Astrophysique, CEA Saclay, 91191, Gif-sur- Yvette, France\\
$^{8}$Central Astronomical Observatory of RAS, Pulkovskoye Chaussee 65/1, 196140, St. Petersburg, Russia\\
$^{9}$Department of Astrophysics, Astronomy \& Mechanics, School of Physics, Aristotle University of Thessaloniki
}

\date{Accepted 2021 July 9. Received 2021 July 5; in original form 2021 April 13.}

\pubyear{2020}

\begin{document}
\label{firstpage}
\pagerange{\pageref{firstpage}--\pageref{lastpage}}
\maketitle

\begin{abstract}
The objective of this paper is to understand the variance of the far-infrared (FIR) spectral energy distribution (SED) of the DustPedia galaxies, and its link with the stellar and dust properties. An interesting aspect of the dust emission is the inferred FIR colours which could inform us about the dust content of galaxies, and how it varies with the physical conditions within galaxies. However, the inherent complexity of dust grains as well as the variety of physical properties depending on dust, hinder our ability to utilise their maximum potential. We use principal component analysis (PCA) to explore new hidden correlations with many relevant physical properties such as the dust luminosity, dust temperature, dust mass, bolometric luminosity, star-formation rate (SFR), stellar mass, specific SFR, dust-to-stellar mass ratio, the fraction of absorbed stellar luminosity by dust ($f_\mathrm{abs}$), and metallicity. We find that 95\% of the variance in our sample can be described by two principal components (PCs). The first component controls the wavelength of the peak of the SED, while the second characterises the width. The physical quantities that correlate better with the coefficients of the first two PCs, and thus control the shape of the FIR SED are: the dust temperature, the dust luminosity, the SFR, and $f_\mathrm{abs}$. Finally, we find a weak tendency for low-metallicity galaxies to have warmer and broader SEDs, while on the other hand  high-metallicity galaxies have FIR SEDs that are colder and narrower.
\end{abstract}




\section{Introduction} \label{sec:intro}

Interstellar dust has been well established as one of the main regulators of several physical and chemical processes in the interstellar medium (ISM) in galaxies. Dust grains are mainly formed through the process of grain growth when a critical metallicity is reached in the ISM at the late stage of galaxy evolution \citep[e.g.][]{Hirashita_2012MNRAS.422.1263H, Draine_2009ASPC..414..453D, Asano_2013EP&S...65..213A, Zhukovska_2014A&A...562A..76Z, Galliano_2018ARA&A..56..673G, Galliano_2021A&A...649A..18G, Granato_2021MNRAS.503..511G}, except at low-metallicity environments ($12 + \log(\mathrm{O}/\mathrm{H}) < 8)$ where the dust condensation in the ejecta either by the stellar winds of Asymptotic Giant Branch (AGB) stars or core-collapse SuperNovae becomes more important \citep[e.g.][]{Liang_2009ApJ...690L..56L, Asano_2013EP&S...65..213A, Galliano_2018ARA&A..56..673G}. Cosmic dust makes its presence known through $(i)$ the reddening of the stellar spectrum at the shortest wavelengths (0.1--1~$\mu$m); $(ii)$ by emitting the absorbed starlight as thermal radiation at longer wavelengths (3--1000~$\mu$m); and $(iii)$ by depleting heavy elements from the ISM. Specifically, in a typical spiral galaxy dust can absorb, on average, up to 25\% of the emitted stellar radiation \citep[e.g.][]{Viaene_2017A&A...599A..64V, Bianchi_2018A&A...620A.112B}, with the absorption fraction rising up to even $\sim87\%$ in the special cases of (ultra) luminous infrared galaxies \citep[(U)LIRGs; e.g.][]{Paspaliaris_2021A&A...649A.137P}. 

Different processes excite dust grains emitting in the infrared (IR) regime. The mid-infrared (MIR) region of the spectrum is powered mostly by stochastic heating, when individual ultraviolet (UV) photons heat up small dust grains to temperatures above the local thermal equilibrium. On the other hand, dust grains that reached a thermal equilibrium with their local radiation field dominate the far-infrared (FIR) and sub-millimetre (submm) regimes. Since these dust grains are in a thermal equilibrium, their emission could be represented with a modified black-body function, which allows us to infer the dust properties of a given galaxy (i.e. the dust mass, and temperature and emissivity index $\beta$). Another way to probe those properties is by looking at the FIR colour indices \citep[e.g.][]{Boselli_2010A&A...518L..61B, Boselli_2012A&A...540A..54B, Cortese_2014MNRAS.440..942C, Smith_2019MNRAS.486.4166S}. The different FIR colours are useful indicators of the various temperature components in the FIR regime, but also of the emissivity index $\beta$ which holds key information about the chemical composition of dust grains in the diffuse ISM. Specifically, $\beta$ determines the opacity $\kappa_\lambda$ of the dust, with a typical value found to be $\sim 2$ for a carbonaceous and silicate grain composition \citep{Draine_1984ApJ...285...89D}. However, the dust emission is subject to many degeneracies such as the one between the dust temperature and the dust emissivity index \citep{Kelly_2012ApJ...752...55K}. 

Moreover, the spectral shape of the FIR emission varies significantly with the galactic environment and the different intrinsic stellar properties of galaxies, for example the star-formation rate (SFR) or the stellar mass. Pinpointing which physical properties drive the spectral shape of the dust emission is crucial, to interpret the large dispersion seen in many FIR colour-colour diagrams. Many studies have tried to find such correlations but usually end up with weak trends and/or large scatter between the stellar parameters inferred by spectral energy distribution (SED) fitting and the FIR colours \citep[e.g.][]{Boselli_2012A&A...540A..54B, Cortese_2014MNRAS.440..942C, Smith_2019MNRAS.486.4166S}. In some cases the large scatter can be attributed to the inconsistent photometry measurements from galaxy to galaxy and from waveband to waveband, incomplete samples etc. Using the peak and width directly from the FIR SED might be slightly easier to interpret their dependence with the physical properties. For example, \citet{Remy_Ruyer_2015A&A...582A.121R} measured the peak and the width of the dust emission in order to derive the relationship of the SED shape with the sSFR and metallicity. However, the main problem with studying the FIR fluxes themselves is that all the bands are strongly correlated. An interesting technique which could be able to offer new clear trends among galaxy intrinsic properties and the shape of the FIR SEDs, and possibly introduce new variables that correlate better with dust properties, is principal component analysis (PCA). PCA is a statistical technique for data visualisation and dimensionality reduction. The principal components by construction they capture the most variance, and so they are more descriptive of the full FIR SED, allowing us to get a better insight regarding the source of dispersion in the FIR SEDs. This method of analysis has been already applied in many astronomical studies: the reduction of dimensionality for SED libraries \citep{Han_2014ApJS..215....2H}, the SED decomposition of the N11 massive star-forming complex in the Large Magellanic Cloud \citep[LMC;][]{Galametz_2016MNRAS.456.1767G}, the investigation of the dependencies among the main physical quantities of the KINGFISH galaxy sample \citep{Hunt_2019A&A...621A..51H}, the spectral classification of local galaxies \citep{Tous_2020MNRAS.495.4135T}, and to explain the variations in the polycyclic aromatic hydrocarbon (PAH) emission in NGC~2023 \citep{Sidhu_2021MNRAS.500..177S}, among many others.

A relevant study and motivator for our paper, is the one by \citet{Safarzadeh_2016ApJ...818...62S}. The authors have used PCA to study how the dust emission of simulated isolated and merging galaxies, across different redshifts, is connected with various global properties. Identifying the best correlations between their PCA results and the physical properties of their galaxy sample, allowed them to predict the shape of the FIR SED. They have found that a double power law in terms of the IR luminosity and dust mass is sufficient to determine the shapes of the galaxy FIR SEDs. Furthermore, they have showed that galaxy sizes do not improve the predictive power of the FIR SEDs, suggesting that the physical size of a galaxy does not prompt the characteristic dust temperature. We closely follow the technique and notation for the PCA, as outlined by \citet{Safarzadeh_2016ApJ...818...62S}, but instead of using the data of simulated galaxies, we apply their method on the DustPedia galaxy sample \citep{Davies_2017PASP..129d4102D}. The DustPedia galaxies are a unique amalgamation of galaxies all observed by \textit{Herschel} Space Observatory, and have been treated in a uniform and consistent way \citep{Clark_2018A&A...609A..37C}. Our goal is to explain the variance in the FIR SEDs and its link with the intrinsic stellar and dust properties. We should make a note here that it is not our intention to predict the FIR emission. We refer to \citet{Dobbels_2020A&A...634A..57D}, who used a machine learning framework to predict the FIR emission and the dust properties, based on the available UV to MIR observations of the DustPedia and \textit{Herschel}-ATLAS \citep[H-ATLAS;][]{Valiante_2016MNRAS.462.3146V} galaxy samples.

This paper is structured as follows. In Sect.~\ref{sec:dataset} we present the dataset used in this study, and in Sect.~\ref{sec:fir_colours} we show several FIR/submm colour-colour relations. Section~\ref{sec:pca} describes the concept behind the use of PCA, while in Sect.~\ref{sec:discussion} we discuss the discovered trends between the PCA results and their relation to other global physical properties such as SFR, dust mass, dust temperature, etc. Our main conclusions are summarised in Sect.~\ref{sec:conclusions}.

\section{The dataset} \label{sec:dataset}

In this paper we delve into the unique and most complete dataset of FIR SEDs in the local Universe, the DustPedia project\footnote{\url{http://dustpedia.astro.noa.gr}} \citep{Davies_2017PASP..129d4102D, Clark_2018A&A...609A..37C}. The database includes imagery and photometric data for 875 local galaxies (distances of $< 40$~Mpc) all observed by \textit{Herschel} Space Observatory \citep{Pilbratt_2010A&A...518L...1P}. Furthermore, infrared data from other missions are available such as \textit{Spitzer} \citep{Werner_2004ApJS..154....1W}, Wide-field Infrared Survey Explorer \citep[WISE;][]{Wright_2010AJ....140.1868W}, InfraRed Astronomical Satellite \citep[IRAS;][]{Neugebauer_1984ApJ...278L...1N}, and \textit{Planck} \citep{Planck_Collaboration_2011A&A...536A...1P}. As for the UV, optical, and near-infrared (NIR) regimes the available data were collected from GALaxy Evolution eXplorer \citep[GALEX;][]{Morrissey_2007ApJS..173..682M}, Sloan Digital Sky Survey \citep[SDSS;][]{York_2000AJ....120.1579Y, Eisenstein_2011AJ....142...72E}, and 2 Micron All-Sky Survey \citep[2MASS;][]{Skrutskie_2006AJ....131.1163S}, respectively. The photometry has been performed for all galaxies by \citet{Clark_2018A&A...609A..37C} in a uniform and consistent way. For details about the photometry pipeline we refer the reader to \citet{Clark_2018A&A...609A..37C}. Then, \citet{Nersesian_2019A&A...624A..80N} fitted the panchromatic SEDs (UV-submm) of 814\footnote{Out of the 875 galaxies, 61 were excluded according to their flagging codes given by \citet{Clark_2018A&A...609A..37C}.} DustPedia galaxies using \texttt{CIGALE}\footnote{\url{https://cigale.lam.fr/}} \citep{Noll_2009A&A...507.1793N, Boquien_2019A&A...622A.103B}, and delivered their global physical properties along with their modelled SEDs. The dust properties were estimated assuming The Heterogeneous Evolution Model for Interstellar Solids\footnote{\url{https://www.ias.u-psud.fr/themis/THEMIS\_model.html}} \citep[\texttt{THEMIS};][]{Jones_2013A&A...558A..62J, Jones_2017A&A...602A..46J, Kohler_2014A&A...565L...9K} dust model. The \texttt{THEMIS} framework uses laboratory measurements to build the synthetic optical properties of amorphous hydrocarbon and silicate materials. So far, \texttt{THEMIS} was qualitatively consistent with many dust observables of the diffuse Galactic ISM, including the observed FUV--NIR extinction and the shape of the IR to millimetre dust thermal emission.

Another characteristic information about the DustPedia sample is that it contains galaxies of various morphological types, i.e. ellipticals, lenticulars, spirals and irregulars, as well as of different orientations, with inclinations ranging from edge-on to face-on. The morphological type is parameterised with the Hubble stage $T$ that ranges from -5 to 10, taken from HyperLEDA\footnote{\url{http://leda.univ-lyon1.fr/}} \citep{Makarov_2014A&A...570A..13M}. We divide our sample into six main morphological classes, E~[-5,~-3.5); S0~[-3.5,~0.5); Sa--Sab~[0.5,~2.5); Sb--Sc~[2.5,~5.5); Scd--Sdm~[5.5,~8.5); Sm--Irr~[8.5,~10]. In some instances within text we refer to galaxies with $T < 0.5$ as early-type galaxies (ETGs), while galaxies with $T \ge 0.5$ are defined as late-type galaxies (LTGs).

\begin{figure}
	\includegraphics[width=\columnwidth]{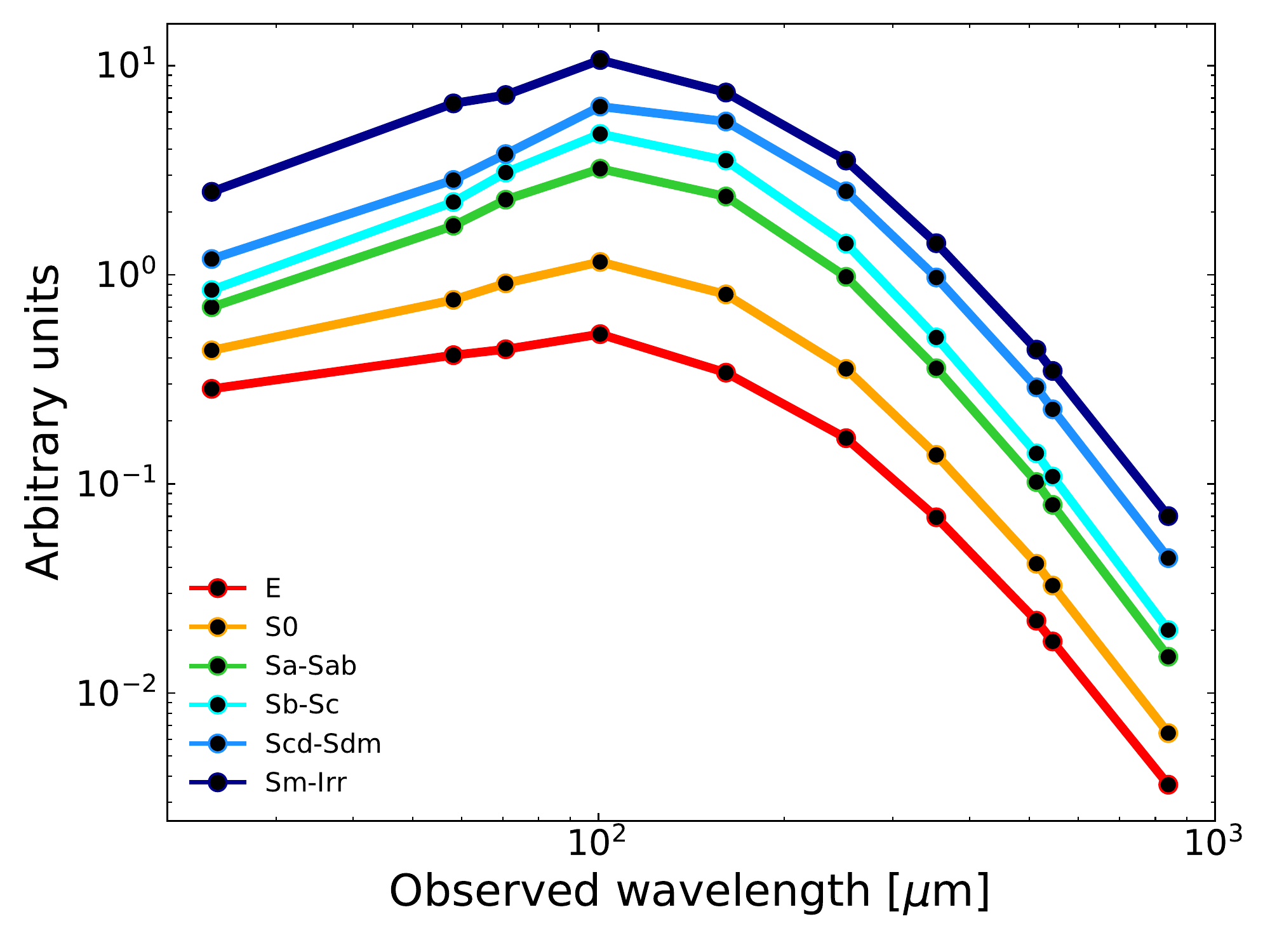}
    \caption{The mean, \textit{Bayesian} FIR SEDs of each morphological group. The FIR SEDs are scaled for visualisation purposes. The \textit{y-axis} is in arbitrary units.}
    \label{fig:bayes_sed}
\end{figure}

For the purposes of this study, we have created a sample of 791 galaxies by excluding 23 peculiar galaxies, to secure that our analysis will not be affected by those objects. Among the excluded galaxies there are 19 harbouring an active galactic nucleus (AGN) and 4 jet-dominated radio galaxies. We use 10~wavebands: MIPS~24~$\mu$m, IRAS~60~$\mu$m, PACS~70,~100,~160~$\mu$m, SPIRE~250,~350,~500~$\mu$m, and \textit{Planck}~550,~850~$\mu$m. In order for PCA to function properly, the complete FIR SED is required. However, it is quite difficult to obtain a complete set of FIR SEDs for all 791 galaxies. Not all galaxies have fluxes in the same set of filters due to heterogeneous data. Subsequently, we chose to use the \textit{Bayesian} flux densities for all bands (not only for the missing values), from the SED models produced with \texttt{CIGALE} and which are available at the DustPedia archive. \texttt{CIGALE} estimates the flux posteriors in likelihood-weighted averages over all SED models. The advantage of utilising the \textit{Bayesian} fluxes, instead of interpolating the observed data, is that they act as some kind of smoothing for bands with large uncertainties; since a realistic library of SEDs is used as a prior \citep[e.g.][]{Dobbels_2020A&A...634A..57D}. Figure~\ref{fig:bayes_sed} shows the mean FIR SEDs of each morphological group. Out of those 10~wavebands, on average, 8 show deviations lower than or equal to 15\% from their observed counterpart, while the IRAS~60~$\mu$m and PACS~100~$\mu$m show the largest deviations (29\% and 17\% respectively)\footnote{A detailed evaluation of how well the flux density in a particular waveband is retrieved by \texttt{CIGALE} is given in \citet{Nersesian_2019A&A...624A..80N}.}. In Appendix~\ref{sec:appA}, we present the Kernel Density Estimation (KDE) of the normalised residuals for each waveband to detect any systematic deviations in the various bands.

\begin{figure*}
	\includegraphics[width=16cm]{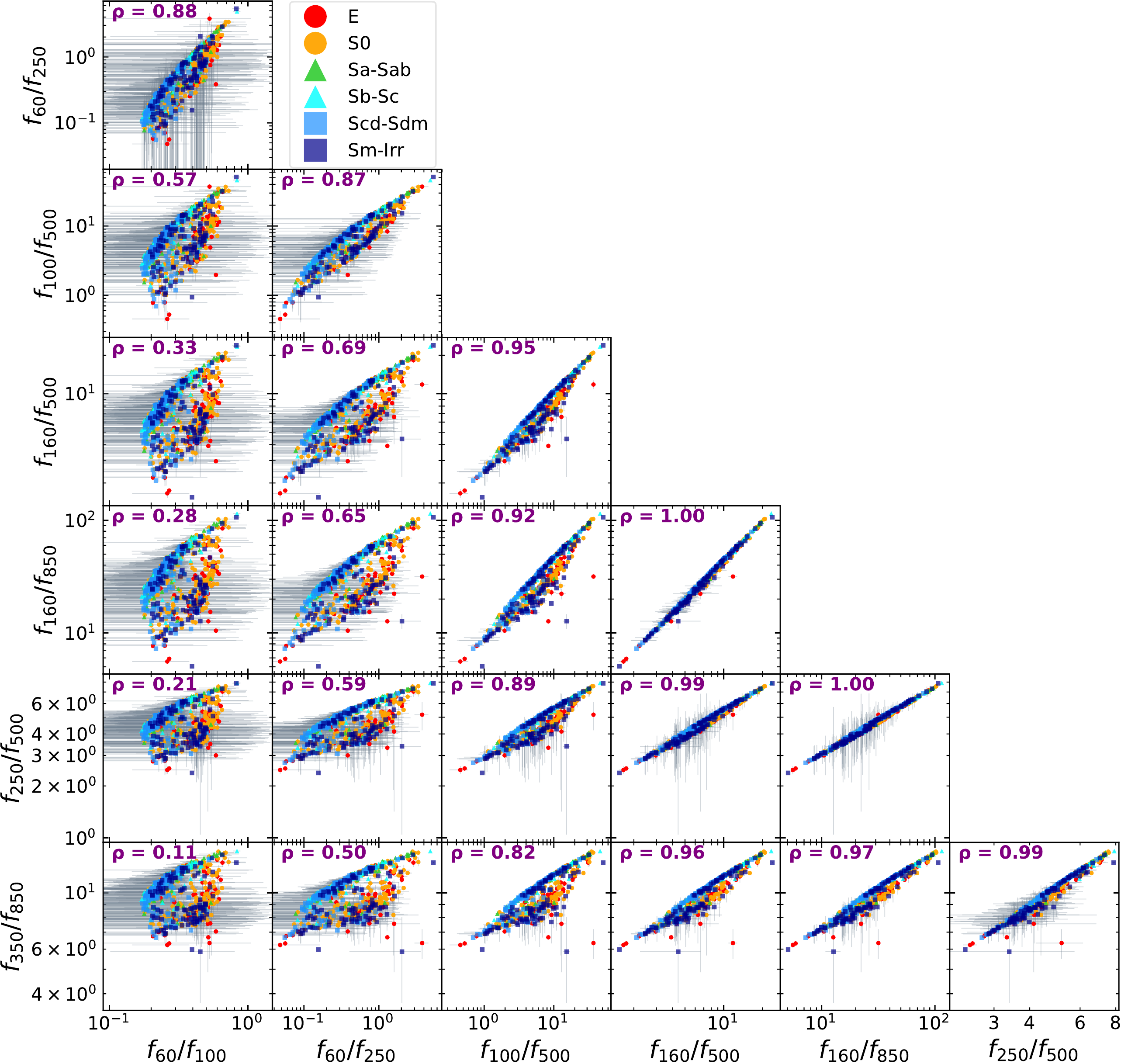}
    \caption{FIR colour–colour diagrams. A FIR colour index is defined as the ratio of the flux densities in two different wavebands. Galaxies are colour-coded according to their morphological type, and divided into six broad morphological groups as indicated in the legend of the figure. The Spearman’s rank correlation coefficient ($\rho$) is given in the \textit{top-left} corner of each panel.}
    \label{fig:fir_colours}
\end{figure*}

\section{The FIR colours} \label{sec:fir_colours}

In this section, we present the flux densities of the DustPedia galaxies in terms of FIR-submm colours. Several of the colour-colour diagrams and their relationships have already been explored for different galaxy samples \citep[e.g.][and many others]{Boselli_2010A&A...518L..61B, Boselli_2012A&A...540A..54B, Bendo_2012MNRAS.419.1833B, Dale_2012ApJ...745...95D, Auld_2013MNRAS.428.1880A, Smith_2019MNRAS.486.4166S}. A colour index is defined as the ratio of the flux densities measured in different FIR bands. Out of all the possible colour-colour combinations, in Fig.~\ref{fig:fir_colours} we show those colour-colour diagrams that are sensitive to the different dust emission components. For example, the $f_{60}/f_{100}$ is more sensitive to the warm dust component mainly heated by young stars, and therefore can be used to trace the star-forming activity in galaxies. A colour index such as the $f_{60}/f_{250}$ is indicative of the average dust temperature, since the involved wavelengths sample the peak of the dust emission. Finally, the colour indices $f_{100}/f_{500}$, $f_{160}/f_{500}$, $f_{160}/f_{850}$, $f_{250}/f_{500}$, and $f_{350}/f_{850}$ sample the Rayleigh-Jeans tail of the dust emission, situating them as good tracers of the coldest dust component ($<15~$K) which carries most of the dust mass in galaxies \citep{Siebenmorgen_1999A&A...351..495S}.

A selection of several FIR colour indices and their in-between relationships are shown in Fig.~\ref{fig:fir_colours}. The different colour-coded markers represent galaxies that fall into six broad morphological groups. We rank the strength of the relationships among the colour indices with the Spearman’s rank correlation coefficient ($\rho$), given in the \textit{top-left} corner of each panel. Looking at the $\rho$ values it is clear that all colour indices are closely related. The FIR colours which cover the closest spectral bands are also those with the tightest relations, for example $f_{160}/f_{850}$ vs $f_{160}/f_{500}$ and $f_{250}/f_{500}$ vs $f_{160}/f_{850}$. We find a weakened relation between the colour indices sensitive to the warm dust component (i.e. $f_{60}/f_{100}$) and the colour indices tracing the emission of the cold dust component ($f_{160}/f_{500}$, $f_{160}/f_{850}$, $f_{250}/f_{500}$ and $f_{350}/f_{850}$). Similar trends have been found for the \textit{Herschel} Reference Survey \citep[HRS][]{Boselli_2010PASP..122..261B}\footnote{Most of the HRS galaxies are included in the DustPedia sample.} by \citet{Boselli_2012A&A...540A..54B} and \citet{Cortese_2014MNRAS.440..942C}, as well as for the JINGLE survey in \citet{Smith_2019MNRAS.486.4166S}. 

Another noticeable trend is an increased dispersion for the colour-colour relations which trace the warm and cold dust component. Yet the scatter for our sample is not as large as described in e.g. \citet{Boselli_2012A&A...540A..54B}, and the reason is that we use the \textit{Bayesian} and not the observed flux densities. The IRAS~60~$\mu$m band ($f_{60}$) is responsible for the large uncertainties in the first two columns of Fig.~\ref{fig:fir_colours}. The galaxies also appear to be clearly separated in two modes according to their morphological type, especially in the first three columns of Fig.~\ref{fig:fir_colours}. The LTGs fall in the "upper" mode suggesting a colder dust component, while the ETGs fall in the "lower" mode. Interestingly, many of the Sm-Irr galaxies fall in the "lower" mode as well, along with the ETGs. This is a clear indication that the spectral shape of the dust emission varies and, therefore, the dust properties significantly change with the evolution state of galaxies. In the following sections we will try to investigate which physical parameters are exactly responsible for the variations in the shape of the FIR SED, by analysing the dust emission with PCA. 

\section{Fitting the FIR SEDs with PCA} \label{sec:pca} 

PCA is a widely used method to explore the variance of large datasets and for creating predictive models. A regular application of this method is to reduce the dimensionality of a large dataset and visualise the relation between different parameters. The principal components are defined to be the eigenvectors of the data's covariance matrix, and they are designed to capture as much variance as possible. The first component captures the most variance, followed by the second, and so forth, revealing the internal structure of the data in a way that best explains the variance in the data.

As already mentioned in Sect.~\ref{sec:dataset}, we make use of the full FIR SEDs for each galaxy as derived with \texttt{CIGALE}. First, we normalise the SEDs with the respective IR luminosity, defined as the integral of the SED over the wavelength range 8-1000~$\mu$m \citep[e.g.][]{Kennicutt_1998ARA&A..36..189K}. Then we apply the PCA using the \texttt{python module sklearn}, which finds the mean SED of our sample and consequently recovers the principal components that could describe the observed variance. The fraction of the variance in the data that can be explained by the different PCs, determines also the rank of each PC. Given a galaxy \textit{j} in the sample, it becomes possible to reconstruct its observed FIR SED through the linear combination of the different PCs:

\begin{equation} \label{eq:pca_sed_recon}
    \lambda F^\mathrm{norm}_{\lambda, j} = \big\langle \lambda F^\mathrm{norm}_{\lambda} \big\rangle + \sum_{i=1}^{N} C_{i,j} \times \mathrm{PC}_{i},
\end{equation}

\noindent
where $\big\langle \lambda F^\mathrm{norm}_{\lambda} \big\rangle$ is the mean FIR SED, $N$ is the number of PCs which is the same as the number of FIR bands that were used, PC$_{i}$ is the $i$th PC, and $C_{i,j}$ is the coefficient of the $i$th PC for the $j$th galaxy. After the identification of the main PCs, we can examine how their coefficients are related to various global physical quantities retrieved from a panchromatic SED fitting. 

\begin{figure}
	\includegraphics[width=\columnwidth]{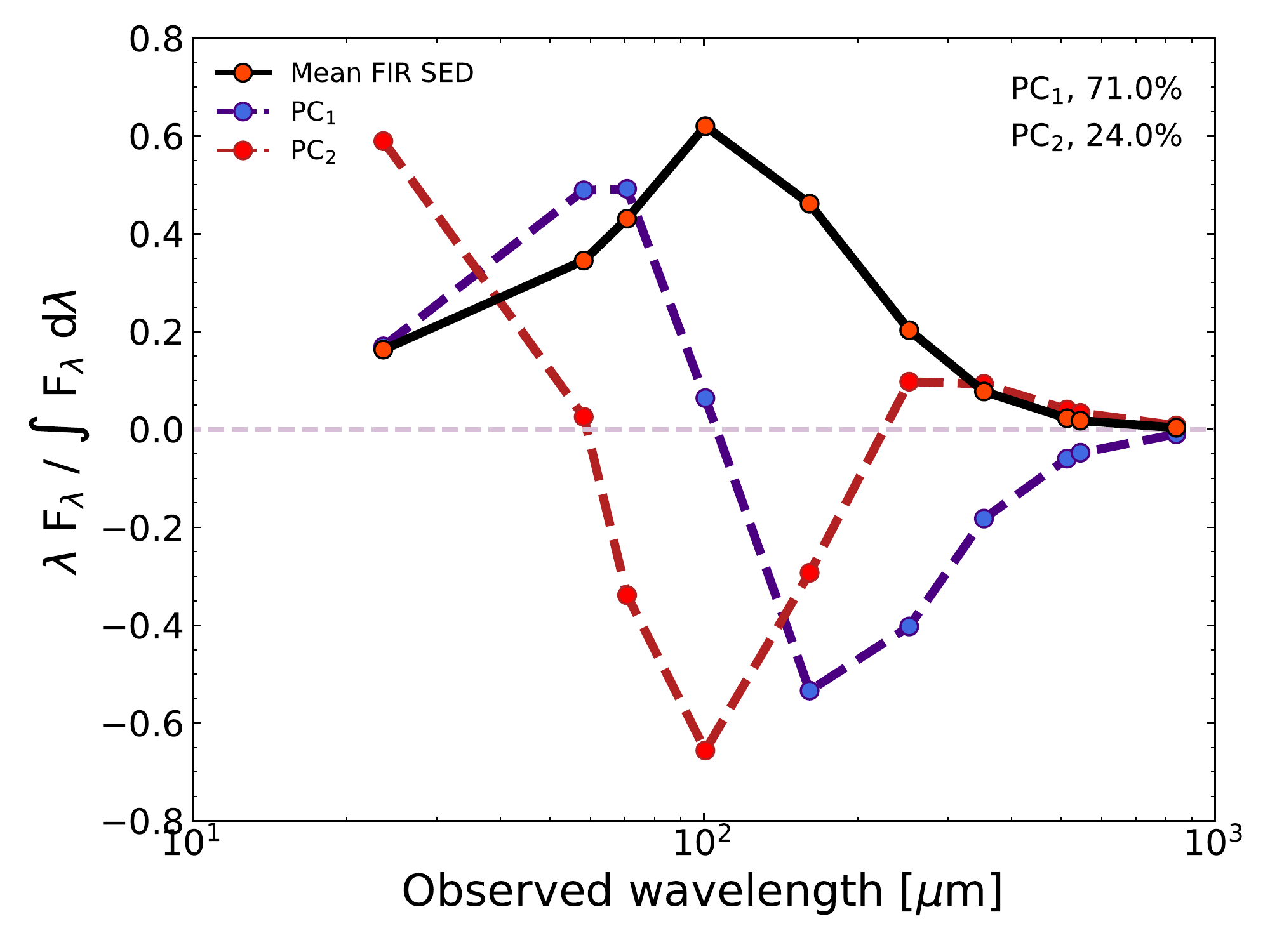}
    \caption{The mean FIR SED of our sample and its first two principal components. The solid black line corresponds to the mean FIR SED, normalised with the IR luminosity. The dashed blue line corresponds to the first PC and the dashed red line to the second PC. The fraction of the variance in the data that is explained by each PC is indicated in the legend. Together, the first two PCs account for 95\% of the total variance of the data in our sample.}
    \label{fig:sed_pcs}
\end{figure}

Figure~\ref{fig:sed_pcs} shows the mean FIR SED of our galaxy sample and the main two principal components which can account for $95\%$ of the variance. Specifically, the first principal component explains $71\%$ of the variance while the second component explains $24\%$. The decomposition of the FIR SED into PCs is in excellent agreement with the results reported by \citet{Safarzadeh_2016ApJ...818...62S}. In particular, \citet{Safarzadeh_2016ApJ...818...62S} have found that 97\% of the variance in their SED sample can be captured by the first two PCs (PC$_1$, 75\%; and PC$_2$, 22\%). The general approach to reconstruct the FIR SED of a given galaxy is to multiply each PC with a coefficient unique for each galaxy and then add it to the mean SED of the sample (see Eq.~\ref{eq:pca_sed_recon}). The coefficients of each PC can take either positive or negative values and can determine how the FIR SED varies from the mean SED. We refer to the coefficients of PC$_1$ as $C_1$ and of PC$_2$ as $C_2$.

\begin{figure}
	\includegraphics[width=\columnwidth]{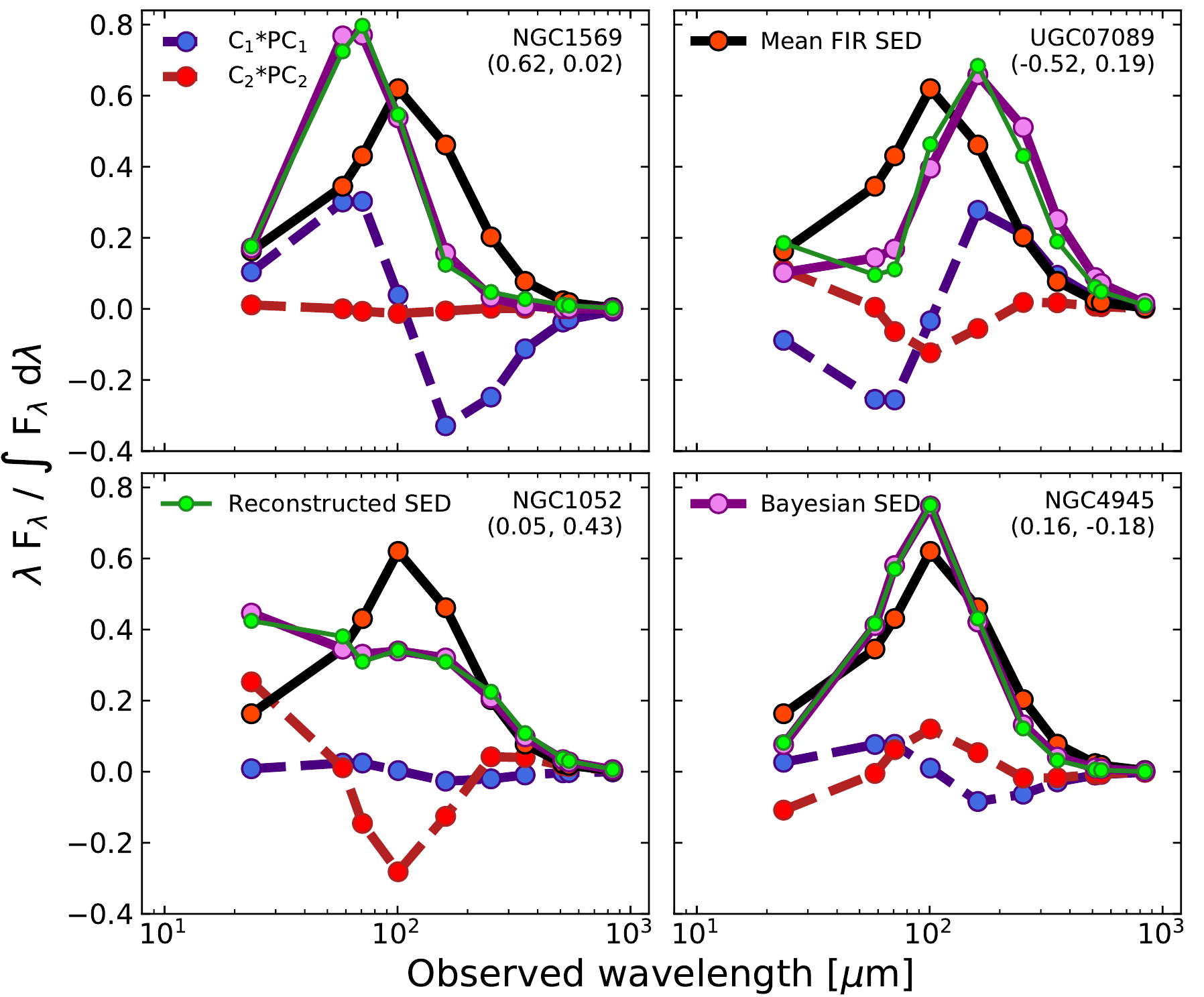}
    \caption{Examples of reconstructed FIR SEDs with the first two principal components. The solid black line corresponds to the mean FIR SED, normalised with the IR luminosity. The dashed blue line corresponds to the first PC and the dashed red line to the second PC. The solid purple line is the observed SED using the \textit{Bayesian} fluxes from \texttt{CIGALE}, while the solid green line shows the SED, reconstructed out of the first two PCs (see Eq.~\ref{eq:pca_sed_recon}). At the \textit{top-right} corner of each panel we show the unique pair of coefficients ($C_1$, $C_2$), for the corresponding galaxy. Any differences between the reconstructed and the observed SEDs are due to the omission of the remaining PCs.}
    \label{fig:sed_pca_examples}
\end{figure}

\begin{figure}
	\includegraphics[width=\columnwidth]{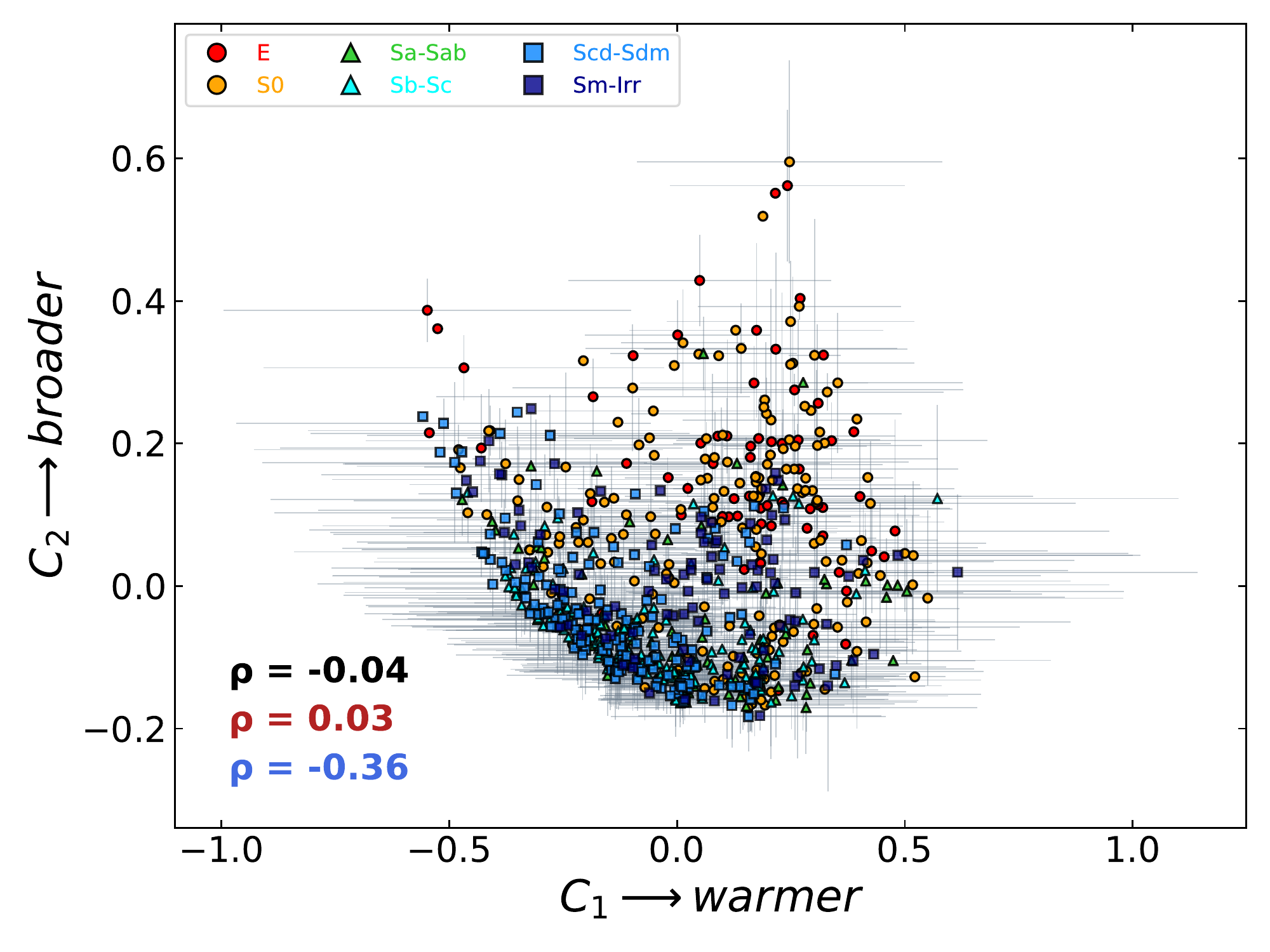}
    \caption{Relation between $C_1$ and $C_2$. Each point represents a galaxy with the different markers and colours indicating the morphological group that they fall into. The Spearman’s rank correlation coefficient ($\rho$) is given in the \textit{bottom-left} corner for the full sample (black font), only the ETGs (red font), and only the LTGs (blue font).}
    \label{fig:c1_c2_hsg}
\end{figure}

\begin{figure*}
	\includegraphics[width=\textwidth]{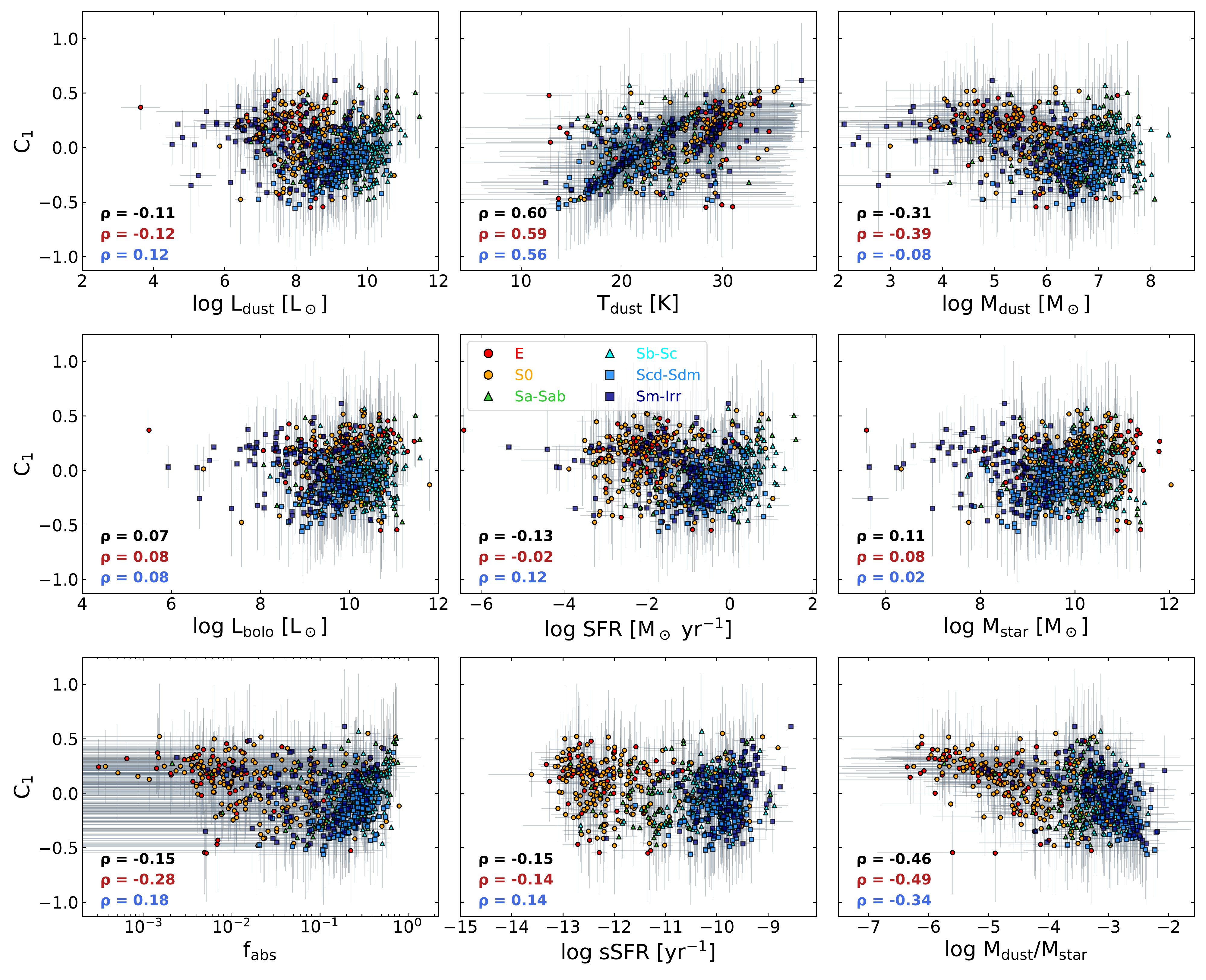}
    \caption{Dependence of the coefficients of PC$_1$ with nine global physical parameters of the galaxies in our sample. \textit{Top-row}: The dust luminosity, the dust temperature, and the dust mass. \textit{Middle-row}: The bolometric luminosity, the star-formation rate, and the stellar mass. \textit{Bottom-row}: The fraction of bolometric luminosity absorbed by dust, the specific star-formation rate, and the dust-to-stellar mass ratio. Galaxies of different morphological groups are indicated with different colours and markers as shown in the legend of the figure. The Spearman’s rank correlation coefficient ($\rho$) is given in the \textit{bottom-left} corner for the full sample (black font), only the ETGs (red font), and only the LTGs (blue font).}
    \label{fig:pca_phys_param_c1}
\end{figure*}

\begin{figure*}
	\includegraphics[width=\textwidth]{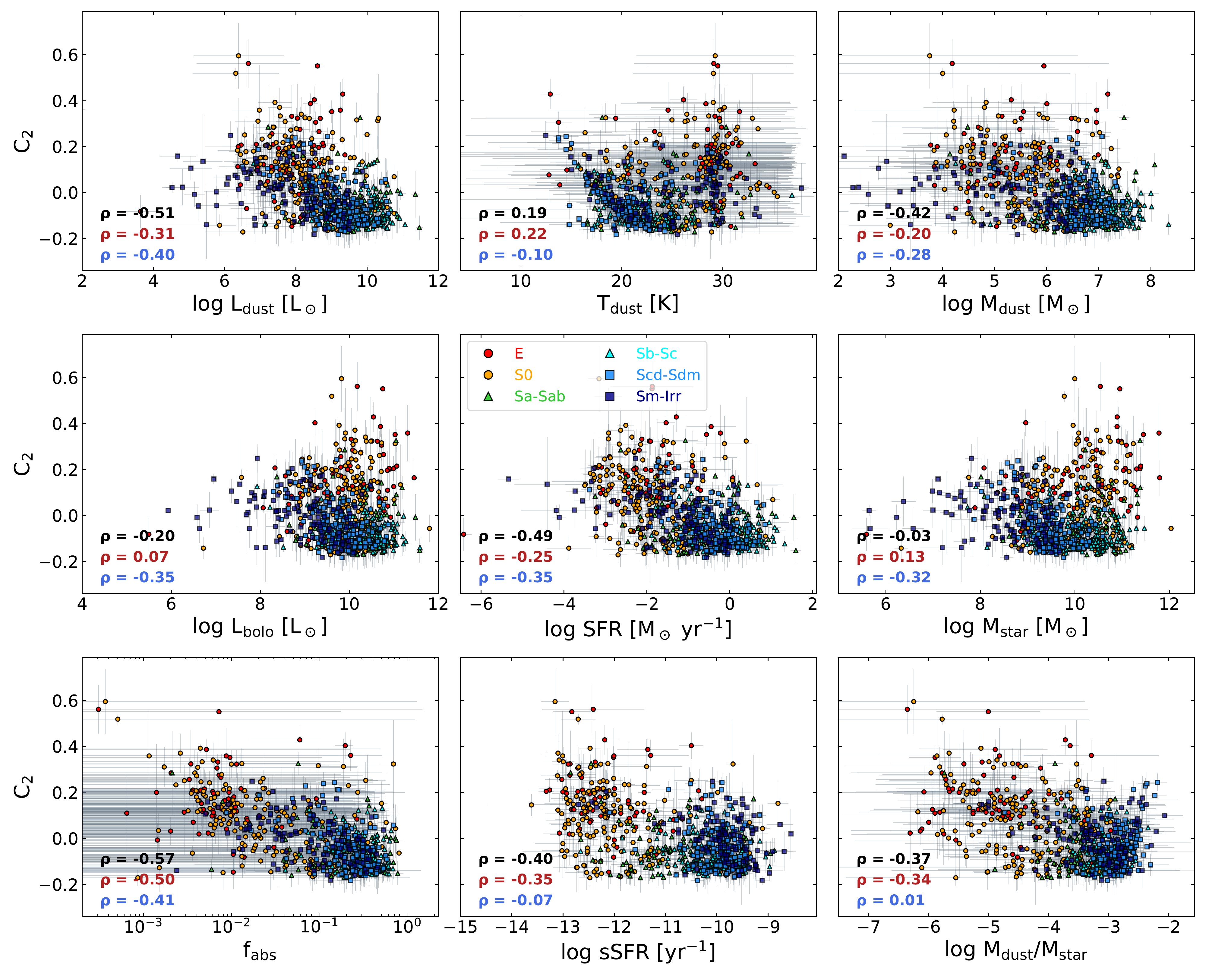}
    \caption{Same as Fig.~\ref{fig:pca_phys_param_c1} but for the coefficients of PC$_2$.}
    \label{fig:pca_phys_param_c2}
\end{figure*}

For the interpretation of the two PCs we look at Fig.~\ref{fig:sed_pca_examples}, where four SED examples are shown. The \textit{top-row} shows the effect of PC$_1$ and the \textit{bottom-row} of PC$_2$. If PC$_1$ was added with a positive coefficient to the mean FIR SED then the peak of the dust emission shifts towards shorter wavelengths (i.e. warmer dust component; \textit{top-left} panel), whilst if PC$_1$ was added with a negative coefficient the peak shifts towards longer wavelengths (i.e. colder dust component; \textit{top-right} panel). Similarly, if PC$_2$ was added with a positive coefficient to the mean FIR SED then the peak of the emission becomes broader (\textit{bottom-left} panel), while if PC$_2$ was added with a negative coefficient the FIR SED becomes narrower (\textit{bottom-right} panel). Therefore, we can infer that PC$_1$ affects the wavelength at which the dust emission has its peak, whereas PC$_2$ controls the width of the FIR SED.

Another interesting result is presented in Fig.~\ref{fig:c1_c2_hsg}, which depicts the relationship between $C_1$ and $C_2$ or in other words the relationship between the temperature proxy of the SED and its broadness, respectively. Each point represents a DustPedia galaxy and it is colour-coded according to each galaxy's morphological type. The Spearman’s rank correlation coefficient ($\rho$) is given as well. We estimated the uncertainties of the coefficients with a bootstrap technique, by re-sampling the SEDs according to their uncertainty distribution. At first glance, the two quantities seem uncorrelated ($\rho=-0.04$ if we consider the full sample). However, if we include the information of the morphology for each galaxy we draw another picture. A mild anti-correlation exists between $C_1$ and $C_2$, with $\rho=-0.36$ if we consider only the LTGs. In particular, the anti-correlation is more tight for galaxies of type Sa--Sab, Sb--Sc, and Scd--Sdm ($\rho=-0.47$). Evidently, these galaxies have in general colder and narrower FIR SEDs. In contrast, the E, S0, and many Sm--Irr galaxies, although they showcase a higher degree of dispersion, they mostly occupy the parameter space which indicates broader FIR SEDs and a warmer dust component. These results could explain the split of the galaxy populations in two modes, visible in the first three columns of Fig.~\ref{fig:fir_colours}. Finally, most of the ETGs reside above the parameter space occupied by LTGs. Thus, a potential application of  Fig.~\ref{fig:c1_c2_hsg} is to probe the morphological characteristics of unresolved galaxies in large surveys, like e.g. GAMA \citep{Driver_2009A&G....50e..12D} or COSMOS \citep{Scoville_2007ApJS..172....1S}.

\section{Discussion} \label{sec:discussion}

In this section, we focus on the correlations between the coefficients of PC$_1$ and PC$_2$, which control the peak wavelength and width of the FIR SED for each galaxy, and the various global physical parameters that were retrieved from a panchromatic SED fitting \citep{Nersesian_2019A&A...624A..80N} with \texttt{CIGALE} (see Figs.~\ref{fig:pca_phys_param_c1} and \ref{fig:pca_phys_param_c2}). We specify these parameters as: the dust luminosity ($L_\mathrm{dust}$), defined as the integral of the total SED over the wavelength range 8--1000~$\mu$m; the stellar mass ($M_\mathrm{star}$); the SFR; the specific SFR (sSFR = SFR/$M_\mathrm{star}$); the dust mass ($M_\mathrm{dust}$); the dust-to-stellar mass ratio ($M_\mathrm{dust}/M_\mathrm{star}$); the dust temperature ($T_\mathrm{dust}$); the bolometric luminosity ($L_\mathrm{bolo}$); and the fraction of absorbed stellar luminosity by dust ($f_\mathrm{abs} = L_\mathrm{dust}/L_\mathrm{bolo}$). An interesting result that is quite evident from both figures is the clear separation of the two main galaxy populations, i.e. LTGs (blue and green points) and ETGs (red and orange points). Therefore, it is more constructive to discuss our results by distinguishing the galaxies in our sample as ETGs and LTGs.

Starting with Fig.~\ref{fig:pca_phys_param_c1} and the \textit{top-row}, we notice a very weak dependence of $L_\mathrm{dust}$ with $C_1$ for the whole sample ($\rho=-0.11$). The strength of the relation does not change if we separate the LTGs from the ETGs. We find a weak positive trend between $L_\mathrm{dust}$ and $C_1$ for the LTGs ($\rho=0.12$); and a negative trend for the ETGs ($\rho=-0.12$). A stronger trend can be achieved for the LTGs ($\rho=0.30$), by excluding the Sm--Irr types which have a large dispersion. The strongest dependence of $C_1$ is with $T_\mathrm{dust}$ (full sample $\rho=0.60$). A clear separation exists for LTGs and ETGs, with a tight increasing trend for both galaxy populations. ETGs have, in general, higher dust temperatures than the LTGs as already found by \citet{Nersesian_2019A&A...624A..80N} for the DustPedia galaxy sample, and also shown for the KINGFISH \citep{Skibba_2011ApJ...738...89S}, HRS \citep{Smith_2012ApJ...748..123S}, and HeViCS \citep{Auld_2013MNRAS.428.1880A} projects. Many of the Sm--Irr galaxies occupy the same parameter space with the ETGs, but the mechanisms that drive the dust temperature are different. In the case of ETGs, both the hardness of the radiation field of an older stellar population, as well as the X-ray emitting hot gas are responsible for the warmer dust \citep[e.g.][]{Natale_2010ApJ...725..955N}, whereas in the case of Sm--Irr galaxies the high sSFR (high SFR and low $M_\mathrm{star}$) is the main driver of the dust temperature \citep[e.g.][]{Remy_Ruyer_2015A&A...582A.121R}. Finally, we notice a negative relation between $C_1$ and $M_\mathrm{dust}$ ($\rho=-0.31$), with ETGs having a moderate negative trend ($\rho=-0.39$), whereas the correlation for the LTGs is negligible ($\rho=-0.08$). 

The \textit{middle-row} of Fig.~\ref{fig:pca_phys_param_c1} depicts the relation of $C_1$ with the intrinsic stellar properties of our sample. Looking at the individual middle panels, we notice that $C_1$ does not exhibit any strong trends with the stellar properties. The $\rho$ values somewhat improve for the LTGs if we exclude the Sm--Irr types ($L_\mathrm{bolo}$, $\rho=0.25$; SFR, $\rho=0.27$; $M_\mathrm{star}$, $\rho=0.19$), but still the trends remain weak. Although the positive trend of $C_1$ with $M_\mathrm{star}$ is pretty faint, it might still indicate the impact of the stellar mass on the peak wavelength of the FIR SED. For example, in ETGs the stellar mass concentration and the X-ray luminosity from hot halo gas increases with increasing stellar mass, resulting in the warmer SEDs. Finally, in the \textit{bottom-row} of Fig.~\ref{fig:pca_phys_param_c1}, we show the dependence of $C_1$ with three different ratios. Immediately, we notice that all three ratios follow a similar trend with $C_1$ (and $C2$; see \textit{bottom-row} of Fig.~\ref{fig:pca_phys_param_c2}). This is an expected behaviour, since the three ratios correlate with each other. \citet{Bianchi_2018A&A...620A.112B} and \citet{Trcka_2020MNRAS.494.2823T} showed a strong correlation of sSFR with $f_\mathrm{abs}$ and the dust-to-stellar mass ratio, respectively. A correlation between $f_\mathrm{abs}$ and the dust-to-stellar mass ratio is also expected, since the dust mass and stellar mass are calculated from their respective luminosities.

Furthermore, a clear separation between the two galaxy populations is visible. Out of the three quantities, the dust-to-stellar mass ratio has the stronger trend with $C_1$ (full sample $\rho=-0.46$). The trend becomes even stronger for the ETGs ($\rho=-0.49$). This anti-correlation suggests that galaxies with a lower dust-to-stellar mass ratio tend to have warmer dust SEDs. In the \textit{bottom-left} panel, we see that a faint decreasing trend exists between $C_1$ and $f_\mathrm{abs}$ ($\rho=0.33$). The $C_1$ coefficients of the LTGs exhibit a weak correlation with $f_\mathrm{abs}$ ($\rho=0.18$ or $\rho=0.33$ if we exclude the Sm--Irr types), oppositely to the ETGs which show a mild anti-correlation ($\rho=-0.28$) yet with a larger dispersion. Similarly, and despite the distinct clustering of the two galaxy populations in the \textit{bottom-middle} panel, the relation between the sSFR and $C_1$ is quite weak ($\rho=-0.15$). Even by looking at the two galaxy populations separately, we get a faint correlation with sSFR ($\rho=0.14$) for the LTGs, and a weak anti-correlation ($\rho=-0.14$) for the ETGs. From the \textit{bottom-row} of Fig.~\ref{fig:pca_phys_param_c1}, there is evidence that the peak wavelength of the dust emission is affected, to a various degree, by the stellar mass concentration in the case of ETGs (low dust-to-stellar ratio, $f_\mathrm{abs}$, and sSFR), and by the process of star-formation in the case of LTGs (high dust-to-stellar ratio, $f_\mathrm{abs}$, and sSFR).  

By shifting the attention to Fig.~\ref{fig:pca_phys_param_c2} and if we consider the full sample, $C_2$ follows an anti-correlation trend with most of the depicted quantities. Among the tightest relations are the one with the SFR ($\rho=-0.49$), and the dust luminosity ($\rho=-0.51$). Yet the strongest anti-correlation that is visible, is with $f_\mathrm{abs}$ ($\rho=-0.57$). Highly star-forming galaxies contain more dust, and therefore higher fraction of the diffuse stellar radiation field is absorbed. Moreover, $C_2$ has a moderate negative trend with the dust-to-stellar mass ratio ($\rho=-0.37$), the sSFR ($\rho=-0.40$), and with $M_\mathrm{dust}$ ($\rho=-0.42$). Actually, the relation between the sSFR and the width of the dust SED is in agreement with the results of \citet{Remy_Ruyer_2015A&A...582A.121R}, who showed that the sSFR plays a significant role shaping the dust SED. A mild negative trend exists also with $L_\mathrm{bolo}$ and $M_\mathrm{star}$ but only in the case of LTGs. Finally, we note that for ETGs a weak correlation exists with $T_\mathrm{dust}$ ($\rho=0.22$), and a weak anti-correlation for the LTGs ($\rho=-0.10$ or $\rho=-0.20$ if we exclude the Sm--Irr types). On the one hand, active spiral galaxies possess more diffuse dust, but they also have dust in the star-forming regions where starlight will be efficiently absorbed, resulting in a wider range of temperatures and subsequently to a broader SED. As the effective dust temperature of LTGs increases the SED becomes narrower. This is purely driven by black body behaviour, since warmer black bodies have a more peaked spectrum. On the other hand, the dust in ETGs is concentrated mostly in their central regions, exposed to the extremely dense radiation field of old stars resulting in a FIR SED with a flatter peak. From these results we infer that galaxies tend to have narrower FIR SEDs with increased SFR, dust luminosity, and dust mass.  

\begin{figure}
	\includegraphics[width=\columnwidth]{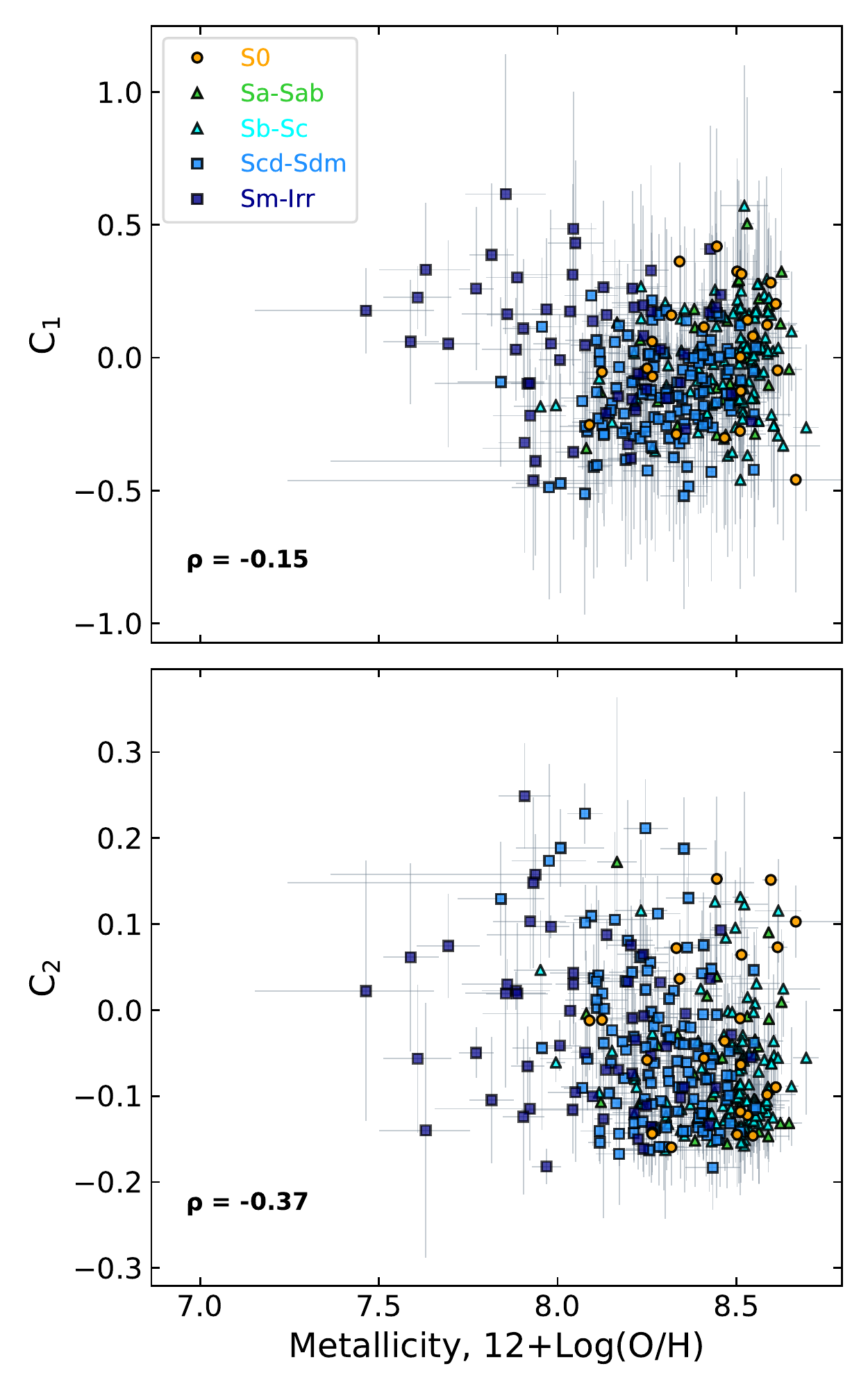}
    \caption{Dependence of the coefficients of PC$_1$ (\textit{top} panel) and PC$_2$ (\textit{bottom} panel) with metallicity. Galaxies of different morphological groups are indicated with different colours and markers as shown in the legend of the figure. The Spearman’s rank correlation coefficient ($\rho$) is given in the \textit{bottom-left} corner.}
    \label{fig:metallicity_c1_c2}
\end{figure}

In addition, we have information about the global metallicities for a fraction of DustPedia galaxies. These metallicities were derived by \citet{De_Vis_2019A&A...623A...5D} using six different calibrations. In particular, we use the S calibration of \citet{Pilyugin_2016MNRAS.457.3678P} as it is the most reliable at low-metallicity \citep{De_Vis_2017MNRAS.464.4680D}. Out of the 791 galaxies in our sample, only 355 galaxies have a metallicity measurement. Figure~\ref{fig:metallicity_c1_c2} depicts the coefficients of the first two PCs as a function of metallicity. For $C_1$ we find a weak negative trend ($\rho=-0.15$), and a mild anti-correlation for $C_2$ ($\rho=-0.37$). We should warn the reader that the correlations presented here may suffer from a bias effect, since we miss the metallicity information for more than half of the galaxies in our sample, and especially for the majority of the ETGs. 

Focusing on the \textit{top} panel of Fig.~\ref{fig:metallicity_c1_c2}, we notice that there is a weak tendency for low-metallicity galaxies to have higher $C_1$ values (i.e. warmer effective dust temperature). Most of the low-metallicity galaxies are Sm--Irr types with positive $C_1$ values, indicating that the peak of their dust emission is shifted towards shorter wavelengths. These results are in good agreement with previous studies, which have demonstrated that low-metallicity galaxies have, in general, warmer dust \citep[e.g.][]{Helou_1985ApJ...298L...7H, Melisse_1994A&A...285...51M, Remy_Ruyer_2014A&A...563A..31R, Remy_Ruyer_2015A&A...582A.121R}. Sm--Irr galaxies are characterised by their clumpy ISM, low stellar mass, and high star formation activity (i.e. high sSFR, see also Fig.~\ref{fig:pca_phys_param_c1}), which is responsible for the warmer dust temperatures. As galaxies evolve and increase in metallicity, their FIR SEDs become colder and narrower as it is shown in the \textit{bottom} panel of Fig.~\ref{fig:metallicity_c1_c2}. Yet the scatter in both panels is large, with several high-metallicity galaxies residing in the parameter space indicative of a hot and broad FIR SED, and vice versa. Our results are consistent with those of \citet{Remy_Ruyer_2015A&A...582A.121R}, who showed that the dust emission for most of the metal-rich KINGFISH galaxies is characterised by a cold and narrow SED, whereas low-metallicity dwarf galaxies have warm and broad SEDs.   

\section{Conclusions} \label{sec:conclusions}

We performed PCA on a sample of FIR SEDs of 791 DustPedia galaxies. Our main goal was to identify the driving source of variation in the FIR SEDs of nearby galaxies, as well as to determine which global physical parameters affect the SED shape the most. Our main conclusions are the following.

\begin{enumerate}

    \item Through the PCA we find that the first two principal components are sufficient to explain 95\% of the variance in the FIR SEDs of our galaxy sample. The first component is associated with the wavelength of the peak of the SED, while the second controls how narrow or broad is the peak. This result is in excellent agreement with the results of \citet{Safarzadeh_2016ApJ...818...62S} for a sample of simulated galaxies.
    
    \item A strong anti-correlation exists between the coefficients of the first two PCs $C_1$ and $C_2$, if we consider galaxies of type Sa--Sab, Sb--Sc, and Scd--Sdm. The parameter space occupied by these galaxies indicates that they have, in general, colder and narrower FIR SEDs. In contrast, the E, S0, and Sm--Irr galaxies, although they showcase a higher degree of dispersion, mostly occupy the parameter space indicative of broader and warmer FIR SEDs.
    
    \item $C_1$ (i.e. the position of the peak of the dust emission) correlates strongly with $T_\mathrm{dust}$ as expected, since the peak of the dust emission is controlled by the effective dust temperature. $C_1$ also showcases a moderate anti-correlation with the dust-to-stellar mass ratio. If we only consider the LTGs, $C_1$ has a weak positive trend with $L_\mathrm{dust}$, $L_\mathrm{bolo}$, SFR, sSFR, and $f_\mathrm{abs}$, while there is no trend with $M_\mathrm{star}$ nor with $M_\mathrm{dust}$. On the other hand, if we only consider the ETGs, we find a mild anti-correlation of $C_1$ with $M_\mathrm{dust}$, and $f_\mathrm{abs}$, while $C_1$ is weakly related to $L_\mathrm{dust}$, $L_\mathrm{bolo}$, $M_\mathrm{star}$, SFR, and sSFR. We conclude that the peak wavelength of the dust emission is affected primarily by the effective dust temperature, and to some extent, by the stellar mass concentration in the case of ETGs (low dust-to-stellar ratio, $f_\mathrm{abs}$, and sSFR), and by the process of star-formation in the case of LTGs (high dust-to-stellar ratio, $f_\mathrm{abs}$, and sSFR).
    
    \item $C_2$ (i.e. the width of the dust emission) shows mostly an anti-correlation in relation to the various physical quantities. The stronger anti-correlations are with $f_\mathrm{abs}$, $L_\mathrm{dust}$, SFR, $M_\mathrm{dust}$, and the sSFR. From these results we infer that galaxies tend to have narrower FIR SEDs with increased SFR, dust luminosity, and dust mass.
    
    \item We find a weak tendency for low-metallicity galaxies (primarily Sm--Irr galaxies) to have higher $C_1$ values, i.e. warmer effective dust temperature. As galaxies evolve and increase in metallicity, their FIR SEDs become colder and narrower. These results are consistent with those of \citet{Remy_Ruyer_2015A&A...582A.121R}, who showed that the dust emission for most of the metal-rich KINGFISH galaxies is characterised by a cold and narrow SED, whereas low-metallicity dwarf galaxies have warm and broad SEDs.
    
\end{enumerate}

\section*{Acknowledgements}

We would like to thank the anonymous referee for the helpful comments and suggestions. AN, WD, gratefully acknowledge the support of the Research Foundation - Flanders (FWO Vlaanderen). EMX, gratefully acknowledges the support by Greece and the European Union (European Social Fund-ESF) through the Operational Programme "Human Resources Development, Education and Lifelong Learning 2014-2020" in the context of the project “Anatomy of galaxies: their stellar and dust content through cosmic time” (MIS 5052455). DustPedia is a collaborative focused research project supported by the European Union under the Seventh Framework Programme (2007-2013) call (proposal no. 606847). The participating institutions are: Cardiff University, UK; National Observatory of Athens, Greece; Ghent University, Belgium; Universit\'{e} Paris Sud, France; National Institute for Astrophysics, Italy and CEA, France. This research made use of Astropy,\footnote{\url{http://www.astropy.org}} a community-developed core Python package for Astronomy \citep{Astropy_2013A&A...558A..33A, Astropy_2018AJ....156..123A}. 

\section*{Data Availability}

All the photometric data and the physical properties that were used in the analysis of this work, are publicly available at the DustPedia archive \url{http://dustpedia.astro.noa.gr}.




\bibliographystyle{mnras}
\bibliography{References_update} 



\appendix

\section{Deviation between the observed and the fitted \texttt{CIGALE} fluxes} \label{sec:appA}

\begin{figure*}
	\includegraphics[width=\textwidth]{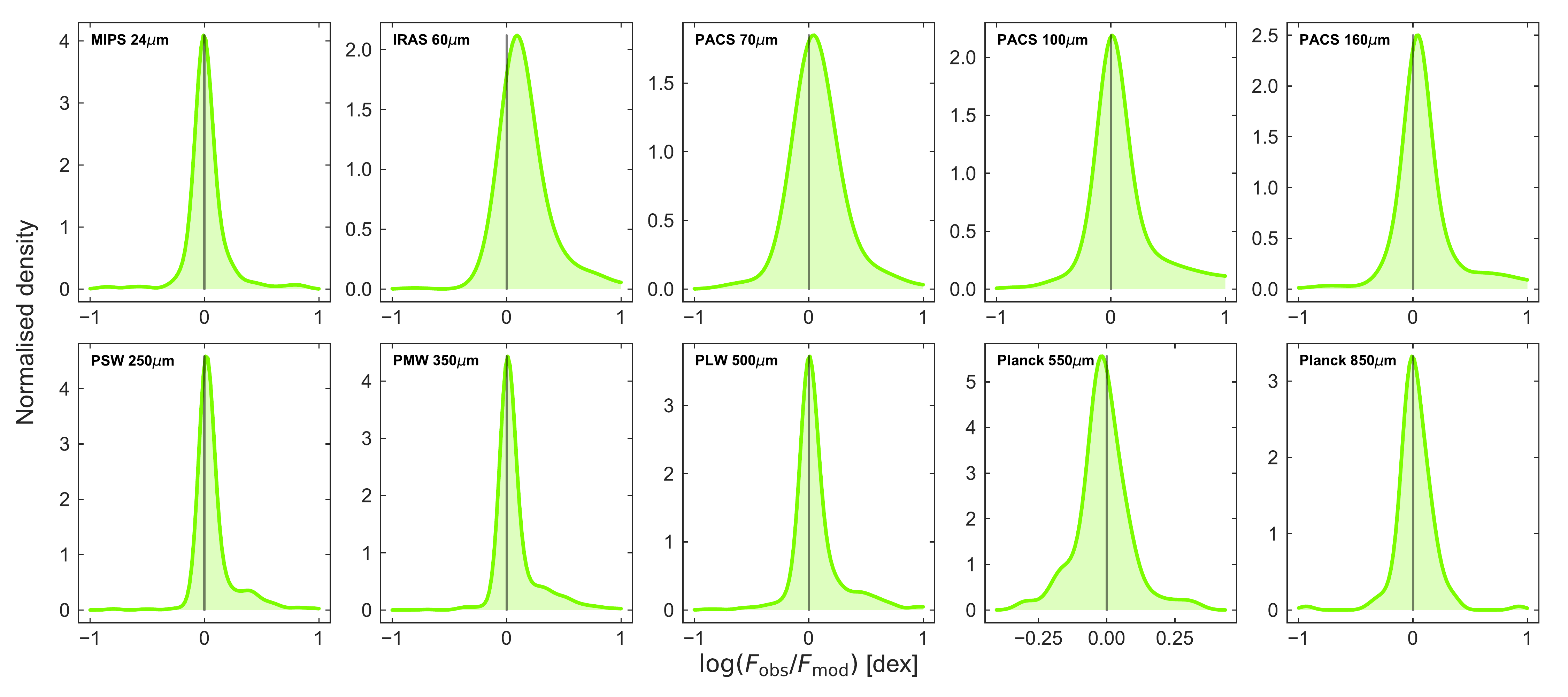}
    \caption{KDE plots of the ratio between the observed fluxes and the best model fluxes, for our galaxy sample. The KDE plot of each waveband is given in each panel. The vertical line declares the zero value.}
    \label{fig:band_comp}
\end{figure*}

In order to validate our choice of using the fitted \texttt{CIGALE} flux densities instead of the observed fluxes in our analysis, we present the Kernel Density Estimation (KDE) of the normalised residuals for each waveband. A systematic effect in the KDE of a given band could help reveal potential weaknesses in our analysis. Figure~\ref{fig:band_comp} reveals that \texttt{CIGALE} is able to reproduce quite accurately the observed fluxes of our galaxy sample. For the majority of the wavebands (7 out of 10), have their KDE of the residuals peaking near $\pm 0.01$~dex and with deviations within $\pm 0.35$~dex. The bands that show a significant deviation are the IRAS~60$\mu$m ($\pm 0.38$~dex), PACS~70$\mu$m ($\pm 0.44$~dex), and PACS~100$\mu$m ($\pm 0.4$~dex). A shift in the peak of the KDE distribution by 0.09~dex, 0.05~dex, and 0.05~dex is also noted for IRAS~60$\mu$m, PACS~70$\mu$m, and PACS~160$\mu$m, respectively. Nevertheless, these deviations are not severe and would not introduce any systematic effects in our current analysis.

\bsp	
\label{lastpage}
\end{document}